\documentclass[a4paper,fleqn,usenatbib]{mnras}
\usepackage{amsmath}
\usepackage{graphicx}
\usepackage{epstopdf}
\usepackage{xcolor}
\newcommand{\hii}    {H\,{\sc{ii}}}
\newcommand{\sii}    {S\,{\sc{ii}}}
\newcommand{\mkm}    {$\mu$m}
\newcommand{\tgas}    {$T_{\rm gas}$}
\newcommand{\ngas}    {$n_{\rm gas}$}

\newcommand{\kms}    {km s$^{-1}$}

\newcommand{\rv}    {R_{\rm V}}
\newcommand{\av}    {A_{\rm V}}

\newcommand{\co}    {$^{13}$CO(3--2)}
\newcommand{\hcop}    {HCO$^+$(4--3)}

\title[H/H$_2$ and C$^+$/C/CO transitions in the Orion Bar] {Merged H/H$_2$ and C$^+$/C/CO transitions in the Orion Bar}

\author[Kirsanova, Wiebe]
{Maria S. Kirsanova$^{1}$\thanks{E-mail: kirsanova@inasan.ru}, Dmitri S. Wiebe$^1$
\\
$^1$Institute of Astronomy, Russian Academy of Sciences, 48 Pyatnitskaya St. 119017, Moscow, Russia\\
}
\begin{document}

\date{Accepted 27 October 2017. Received 27 October 2017; in original form 27 October 2017}

\pagerange{\pageref{firstpage}--\pageref{lastpage}} \pubyear{2017}

\maketitle

\label{firstpage}

\begin{abstract}
High-resolution ALMA images towards the Orion Bar show no discernible offset between the peak of H$_2$ emission in the photodissociation region (PDR) and the \co\ and \hcop\ emission in the molecular region. This implies that positions of H$_2$ and CO dissociation fronts are indistinguishable in the limit of ALMA resolution. We use the chemo-dynamical model MARION to show that the ALMA view of the Orion Bar, namely, no appreciable offset between the \co\ and \hcop\ peaks, merged H$_2$ and CO dissociation fronts, and high intensity of \hcop\ emission, can only be explained by the ongoing propagation of the dissociation fronts through the molecular cloud, coupled to the dust motion driven by the stellar radiation pressure, and are not reproduced in the model where the dissociation fronts are assumed to be stationary. Modelling line intensities, we demonstrate that after the fronts have merged, the angular separation of the \co\ and \hcop\ peaks is indeed unresolvable with the ALMA observations. Our model predictions are consistent with the results of the ALMA observations about the relation of the bright \hcop\ emission to the compressed gas at the border of the PDR in the sense that the theoretical \hcop\ peak does correspond to the gas density enhancement, which naturally appears in the dynamical simulation, and is located near the H$_2$ dissociation front at the illuminated side of the CO dissociation front.

\end{abstract}


\section{Introduction}

Young massive stars vigorously interact with surrounding molecular clouds. Expansion of ionized gas around a young massive star(s) is accompanied by a shock wave propagation into the molecular cloud, as has been shown both analytically \citep{spitzerbook} and with numerical simulations \citep[e.g.,][]{Hosokawa_2005}. But observational evidence of this process in real \hii\ regions is scarce and ambiguous,  \citep{2007A&A...472..835Z,2008MNRAS.388..729K,Gordon_2009,2015ApJ...800..101A}, as it is not easy to prove that a compression wave, accompanied by ionization and dissociation fronts, does {\em move\/} into the molecular cloud.

The Orion Bar is 
an ionized border of a molecular cloud in the South-Eastern part of the Orion Nebula \citep[e.g.,][]{ODell_01}, highly irradiated by far ultraviolet (FUV, $6 < {\rm h}\nu < 13.6$~eV) photons. Regions of this kind are usually referred to as photodissociation regions (PDR). The primary photoionization source of the Orion Nebula is a massive star $\theta^1$C~Ori of the spectral type O7~\citep{2011ApJS..193...24S} or O6.5V~\citep{Tsivilev_2014}. Along with the primary star, other Trapezium stars provide significant contribution to the FUV irradiation of the Orion Bar PDR~\citep{Ferland_2012}. The Orion Bar is seen almost edge-on \citep[see e.g.][]{ODell_01}, so that we can investigate how irradiation from the nearby stars penetrates the molecular cloud and changes its chemical and physical structure \citep[see e.g. studies by][]{Hogerheijde_1995,Jansen_1995,AndreeLabsch_17}.

\citet{Sellgren_90} have found spatial stratification, distinguishing an almost parallel ionization front, 3.3\,\mkm\ PAH emission, and 2\,\mkm\ H$_2$ emission. \citet{Tielens_1993}, using an equilibrium model of a plane-parallel irradiated cloud, have reproduced the $\sim10$\,arcsec offset between the PAH and H$_2$ emission, and also between the H$_2$ and CO emission. The study of \citet{Tielens_1993} is based on previous equilibrium model calculations of \citet{Tielens_1985} and \citet{Tielens_1985_ii}, which are now considered as a classic model of a dense PDR.

However, recent high-resolution ALMA millimetre-band images presented by~\cite{Goicoechea_2016} draw a different picture, in which there is no appreciable offset between a peak of H$_2$ vibrational emission and an edge of \co\ and \hcop\ emission towards the Orion Bar. Locations of the \co\ and \hcop\ emission peaks coincide with the region of brightest H$_2$ emission that appears close to the H$_2$ dissociation front \citep[see Extended Data Figure~2 in][]{Goicoechea_2016}. The location of the \co\ emission edge supposedly coincides with a location of a CO dissociation front. \citet{Goicoechea_2016} were not able to explain the ALMA data with a stationary PDR model, so they suggested that time-dependent PDR modelling could explain both the coincidence of H$_2$, \co\ and \hcop\ emission peaks and high brightness of \hcop\ emission.  Bright \hcop\ emission requires gas number density of about $10^6$~cm$^{-3}$ \citep[see e.g. discussion of the critical density for HCO$^+$ in][based on calculations of \citet{Flower_1999}]{Shirley_2015}, and relative HCO$^+$ abundance greater than $10^{-9}$. 


The goal of the present study is to verify whether time-dependent astrochemical calculations can indeed reproduce the observed features of the Orion Bar, namely, no appreciable offset between the \co\ and \hcop\ emission peaks, merged H$_2$ and CO dissociation fronts, and bright \hcop\ emission. To study the Orion Bar PDR, we utilise a chemo-dynamical model of an expanding \hii\ region, implemented as a computer code MARION. Its development is described in~\cite{Kirsanova_2009}, \cite{Pavlyuchenkov_2013}, \cite{Akimkin_2015}, and \cite{Akimkin_2017}. Results of the chemo-dynamical computation are then passed to the RADEX model~\citep{radexpaper} to calculate intensities of \co\ and \hcop\ lines and to compare them to the results of the ALMA observations presented by \citet{Goicoechea_2016}.

\section{Model calculation}\label{sec:mod}

MARION is a 1D spherically symmetric model of the dynamical and chemical evolution of gas and dust in the vicinity of a massive star. It needs to be adapted for the study of the Orion Bar PDR. We simulate the plane-parallel geometry of the molecular cloud edge by adopting a very large radius for a first inner cell, which is supposed to represent the cloud surface. The cloud thickness is 0.225~pc, and it is divided into 1500 cells, so the spatial resolution is about 30~AU. We take the cloud thickness larger than e.g. 0.1~pc as found by \citet{Salgado_2016} in order to provide a sufficient model time for a shock wave to reach a quasi-stationary solution prior to breaking out of the computational domain.


The radiation field is assumed to be generated by massive stars from the Orion Nebula Cluster. We constructed radiation field using parameters of five massive stars that were defined as `M42 inner' by \citet{Ferland_2012}. The resultant radiation field contains no photons with $\lambda \leq 91.2$~nm to save computational time. In other words, we only consider the PDR, but not the ionized region. We calculated the `M42 inner' radiation field at a distance of 0.22~parsec, which corresponds to the projected distance 111\arcsec\ from the main ionizing star to the Bar according to \citet{Pellegrini_2009} and \citet{AndreeLabsch_17}, and then passed the radiation field to the described above plane-parallel modification of the MARION code. The intensity of the radiation field is $\chi=4.4\cdot10^4$ in units of the Draine field~\citep{Draine_1978} in the wavelength range of $91.2 < \lambda \leq 200$~nm. This $\chi$ value is consistent with the FUV radiation flux near the Orion Bar PDR \citep{Tielens_1985_ii}.

We note that at $\lambda \geq 200$~nm this procedure produces radiation field that is orders of magnitude lower than the `universal' Draine field \citep{Draine_1978} with a long wavelength modification introduced by~\cite{vanDishoeck_1982}, scaled by $\chi=4.4\cdot10^4$. Even if we supplement the radiation field generated by the five massive stars with radiation field of less massive stars from the Trapezium Cluster \citep{Muench_2002}, we still do not obtain the same intensity enhancement in the $\lambda \geq 200$~nm range as the scaled sum of the Draine field and the long wavelength modification. The usage of the radiation field constructed from the real stellar population is important for our calculations due to radiation pressure term in the dynamical equations.

After the upgrade by~\citet{Akimkin_2015} and \citet{Akimkin_2017}, various dust size distributions can be accommodated in the MARION model. Here we use 48 dust size bins and WD16~\citep{Weingartner_2001} grain size distribution for $\rv=5.5$ \citep{Bohlin_1981}. The PAH mass is assumed to account for 4~per cent of the total dust mass. With these assumptions, $A_{\rm V}$ depends linearly on the hydrogen number density with the ratio $A_{\rm V}/N_{\rm H}=5.3\times10^{-22}$~cm$^2$. As in the MARION code dust can move with respect to gas, the value of $A_{\rm V}/N_{\rm H}$ varies with distance and time. To calculate photoreaction rates, we use $\tau(5000\,\mbox{\AA})$ computed for the current dust spatial distribution, when needed (see how $\tau_{\rm V} \equiv \tau(5000\,\mbox{\AA})$ changes with time in Sec.~\ref{sec:Disc}). Heating and cooling processes, which are included in the calculation, are listed in Table~\ref{heatandcool}. Radiation pressure on dust grains and momentum transfer from grains to gas are treated as in \citet{Akimkin_2017}.

\begin{table}
\caption{Heating and cooling processes.}
\label{heatandcool}
\begin{tabular}{@{}ll}
\hline
Process & Reference\\
\hline
\multicolumn{2}{|c|}{\it Heating} \\
Photoionization of C       &\citet{tielensbook}\\
Photoelectric heating      &\citet{Bakes_1994}\\
H$_2$ formation            &\citet{Sternberg_1989}\\
H$_2$ dissociation         &\citet{Stephens_1973}\\ 
FUV H$_2$ pumping          &\citet{Hollenbach_1979}\\
Cosmic rays                &\citet{Jonkheid_2004}\\
\hline
\multicolumn{2}{|c|}{\it Cooling} \\
Hydrogen free-free transition &\citet{Osterbrock_2006}\\

Ly$\alpha$ &\citet{spitzerbook}\\

[O\,{\sc{i}}] $\lambda=63.2 \mu$m       &\citet{drainebook}\\

[O\,{\sc{i}}] $\lambda=6300.3$\AA       &\citet{Tielens_1985}\\

[O\,{\sc{ii}}] $\lambda=3728.8$\AA      &\citet{Hollenbach_1989}\\

[C\,{\sc{ii}}] $\lambda=157.7\mu$m    &\citet{Hollenbach_1989}\\

CO rot\&vib transitions    &\citet{Hollenbach_1979}\\

H$_2$ rot\&vib transitions    &\citet{Neufeld_1993}\\

H$_2$O rot\&vib transitions   &\citet{Hollenbach_1979}\\
OH rot\&vib transitions    &\citet{Hollenbach_1979}\\
\hline
Dust-gas interaction &\citet{tielensbook}\\
\hline
\end{tabular}
\end{table}

Parameters of the model cloud and initial chemical abundances are summarized in Table~\ref{tab:modparam}. Gas number density \ngas\ at the outset of the calculation corresponds to the value estimated by \citet{Goicoechea_2016}. We initially put all carbon and oxygen into CO and O$_2$ and apply `\hii' elemental abundances from the CLOUDY model \citep{Ferland_2013}. Chemical reactions on dust surfaces along with accretion and desorption processes are not considered to save computational time. Gas-phase chemical network is taken from a website of PDR-benchmarking workshop\footnote{\url{http://www.astro.uni-koeln.de/site/pdr-comparison/intro1.htm}}. File {\it rate99\_edited\_incl\_crp.dat}~\citep{Rollig_2007} with cross-sections for most photoreactions is taken from the Leiden photo site~\citep{Heays_2017}. The network is based on the UMIST99 ratefile~\citep{LeTeuff_2000}. A full list of species includes H, H$^+$, H$_2$, H$_2$$^+$, H$_3$$^+$, O, O$^+$, OH$^+$, OH, O$_2$, O$_2$$^+$, H$_2$O, H$_2$O$^+$, H$_3$O$^+$, C, C$^+$, CH, CH$^+$, CH$_2$, CH$_2$$^+$, CH$_3$, CH$_3$$^+$, CH$_4$, CH$_4$$^+$, CH$_5$$^+$, CO, CO$^+$, HCO$^+$, He, He$^+$, e$^-$. Cosmic ray  (CR) ionization rate is taken to be $5\times10^{-17}$\,sec$^{-1}$, but we checked higher values of the ionization rate up to $5\times10^{-16}$\,sec$^{-1}$ as mentioned in Sec.~\ref{sec:Disc}.

\begin{table}
\caption{Model parameters. Initial CO and O$_2$ abundances correspond to the `\hii' elemental abundance set from CLOUDY.}
\label{tab:modparam}
\begin{tabular}{c|c}
\hline
Parameter & Value\\
\hline
$\chi$ at the surface & 4.4$\times10^4$\\
Initial \ngas             & $5\times 10^{4}$\,cm$^{-3}$\\
Initial \tgas             & 10\,K\\
Dust model        & WD16, $R_{\rm v} = 5.5$\\
$x$(H$_2$)          & 0.5  \\
$x$(He)             & 0.1  \\
$x$(CO)             & $3\times 10^{-4}$  \\
$x$(O$_2$)          & $5\times 10^{-5}$  \\
\hline
\end{tabular}
\end{table}

\citet{Rollig_2007} provide eight test cases for the benchmarking of PDR models. Prior to the study of the Orion Bar, we perform all these tests with the corresponding values for element abundances and using MARION in a static mode. Results for the test cases with variable gas and dust temperatures are presented in Appendix~\ref{AppA}.

We use the RADEX model \citep{radexpaper} to calculate \co\ and \hcop\ emission intensities at each time step and in each computational cell. Main parameters of the RADEX calculation are the following. Background radiation temperature is 2.73~K in each computational cell. The line width for calculation of \hcop\ integrated intensity is fixed to 2~\kms\ according to the results by \citet{Goicoechea_2016}. We use LAMDA database \citep{Schoier_2005} in the RADEX calculation and account for collisions of CO and HCO$^+$ with H$_2$ molecules. Collisions of CO with atomic hydrogen \citep{Yang_2013} were also included into the calculations with RADEX. RADEX method number 2 is utilized. To convert spatial offsets to angular offsets and to produce a synthetic observational picture we adopt the distance to the Orion Bar of 414 parsec~\citep{Menten_07} and the line-of-sight extent of the Bar of 0.12~pc~\citep{Werner_1976}. Model distributions of \co\ and \hcop\ emission are convolved with 1\,arcsec ALMA beam at 350~GHz.

\section{H/H$_2$ and C$^+$/C/CO transitions in a stationary calculation}~\label{sec:res1}

\begin{figure*}
\includegraphics[width=5.5cm,angle=270]{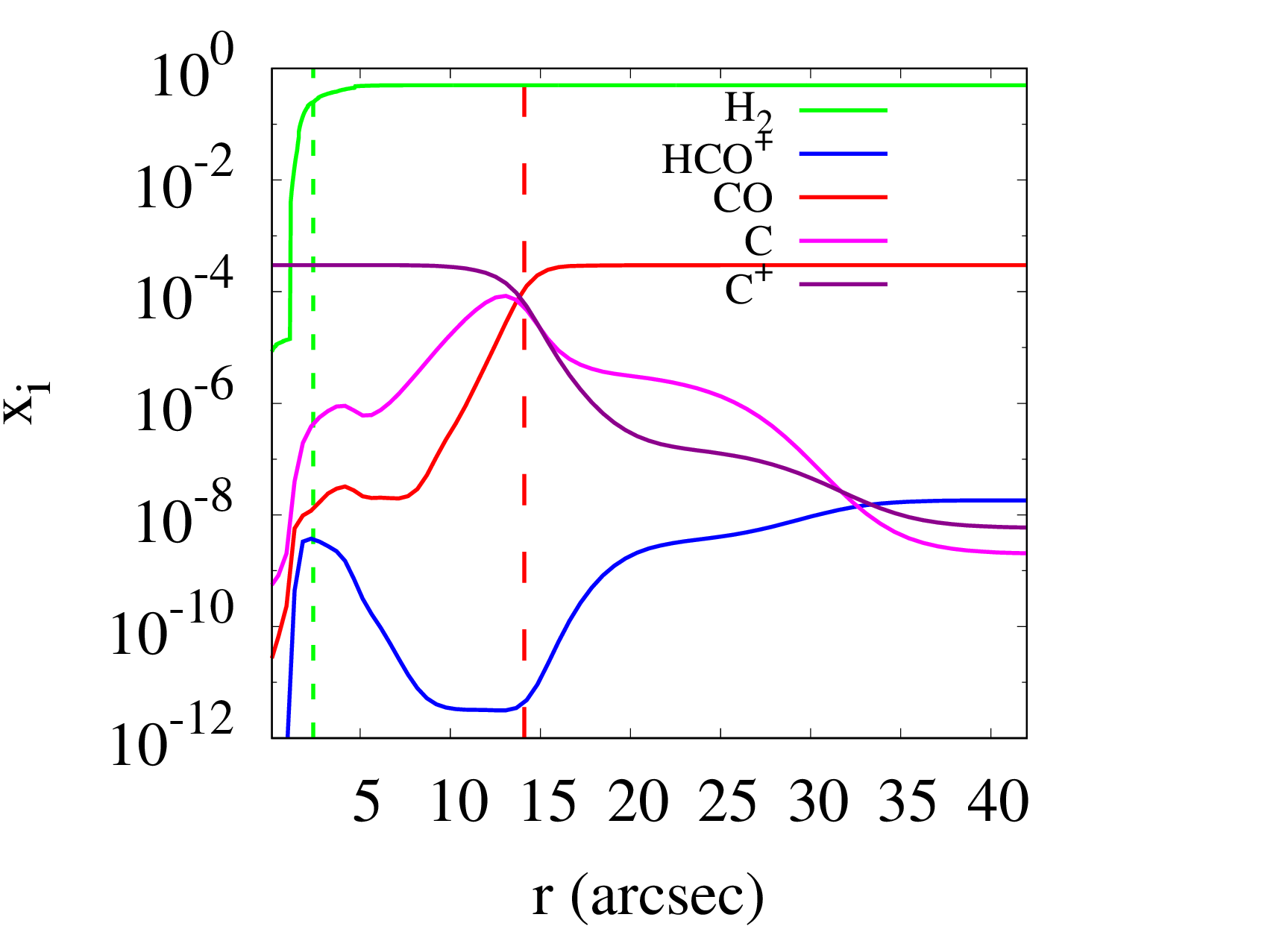}
\includegraphics[width=5.5cm,angle=270]{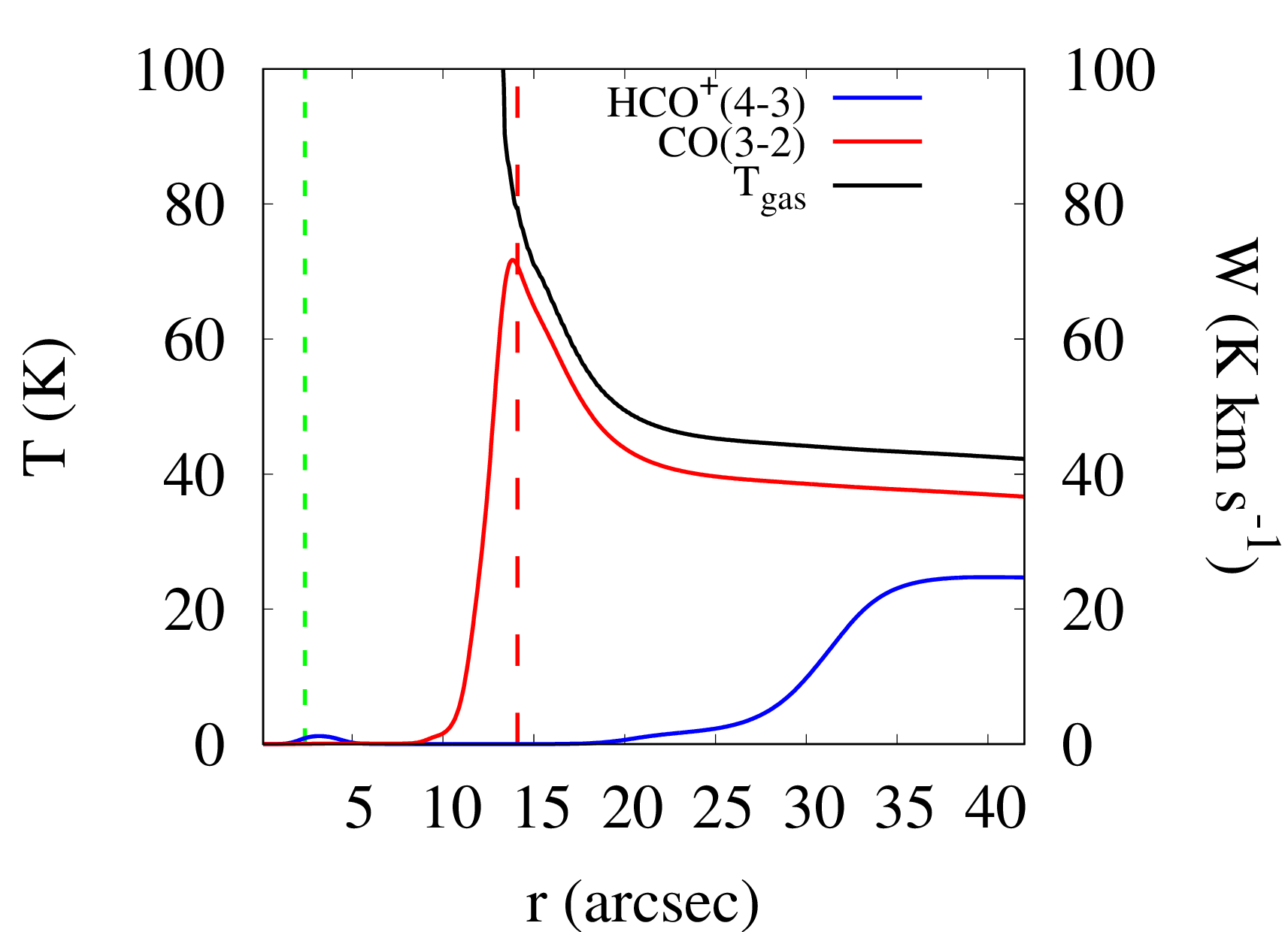}
\caption{Results of the stationary calculation for the Orion Bar. A horizontal axis shows a distance from the inner boundary of the computational domain, $r$. A source of UV photons is supposed to be located beyond the left plot border. Red and green dashed lines show calculated locations of CO and H$_2$ dissociation fronts, respectively.}
\label{fig:general}
\end{figure*}

We begin with calculating a stationary model with parameters described above to compare the computed locations of H$_2$ and CO dissociation fronts to those from the study by \citet{Tielens_1993}. For that purpose, we switch off the dynamical components of the MARION model to calculate a stationary PDR. The run continued for 1.4 Myr to assure stationary conditions (they are actually established much earlier). Results of the calculation are shown in Fig.~\ref{fig:general}. The $r$ value in Fig.~\ref{fig:general},~\ref{fig:timefronts},~\ref{carbon} and~\ref{fronts}  represents a distance from the inner boundary of the computational domain. In the used setup of the MARION model, ionizing photons are not taken into account, so we cannot pinpoint the location of the ionization front. The dissociation front is defined as a location, where an abundance of either H$_2$ or CO drops by a factor of two relative to its initial value given in Table~\ref{tab:modparam}.

In Fig.~\ref{fig:general}, the H$_2$ dissociation front is located right next to the border of the computational domain (on this and subsequent figures it is marked with a vertical green line), while the CO dissociation front (marked with a vertical red line) is shifted deeper into the cloud. The offset between the H$_2$ and CO dissociation fronts is about 12\,arcsec. This result is in agreement with previous equilibrium calculations by \citet{Tielens_1993} and \cite{1994ApJ...422..136T}. The location of the CO dissociation front marks the C$^+$/C/CO transition. Thus, the separation between the H/H$_2$ and C$^+$/C/CO transitions in the stationary model would have been definitely resolved with 1\,arcsec resolution.

Right panel of Fig.~\ref{fig:general} shows results of the RADEX calculation. \citet{Goicoechea_2016} mention that the peak value of the \co\ line intensity is a measure of the gas temperature in molecular clouds. We plot the gas temperature on the right panel of Fig.~\ref{fig:general} and do find it to be very close to the \co\ intensity (within 5~K) everywhere beyond the CO dissociation front. Brightness of the \co\ emission peak is 2.5 times smaller in our results than the value obtained by \citet{Goicoechea_2016}, and the \hcop\ integrated intensity is about 40 times smaller than the value found in the Orion Bar. Thus, the stationary model reproduces neither the merging of the H$_2$, \co\ and \hcop\ emission peaks, seen in the ALMA observations, nor the observed line intensities.

\section{Time-dependent location of H/H$_2$ and C$^+$/C/CO transitions}\label{sec:res2}

\begin{figure*}
\includegraphics[width=4.23cm,angle=270]{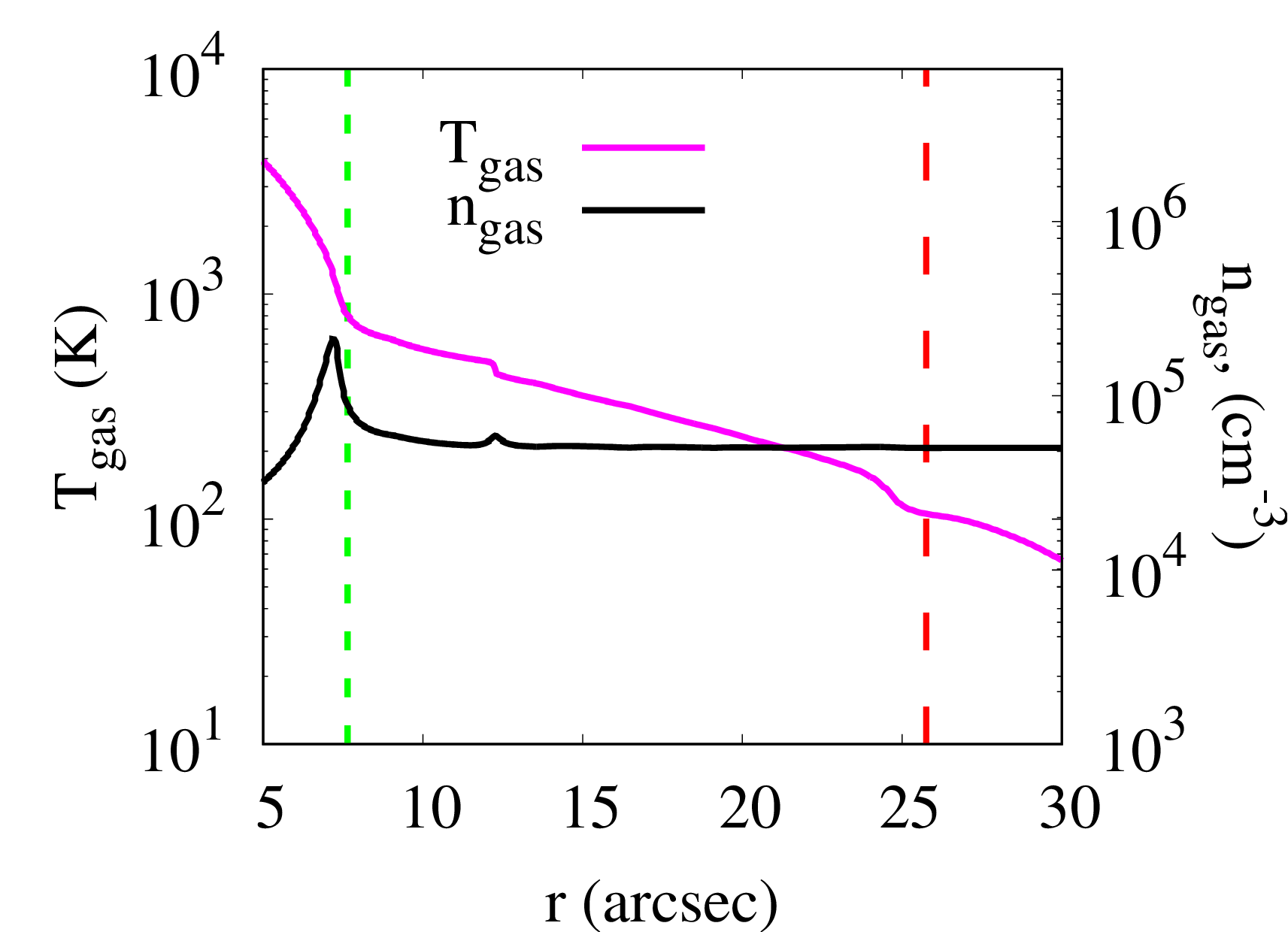}
\includegraphics[width=4.23cm,angle=270]{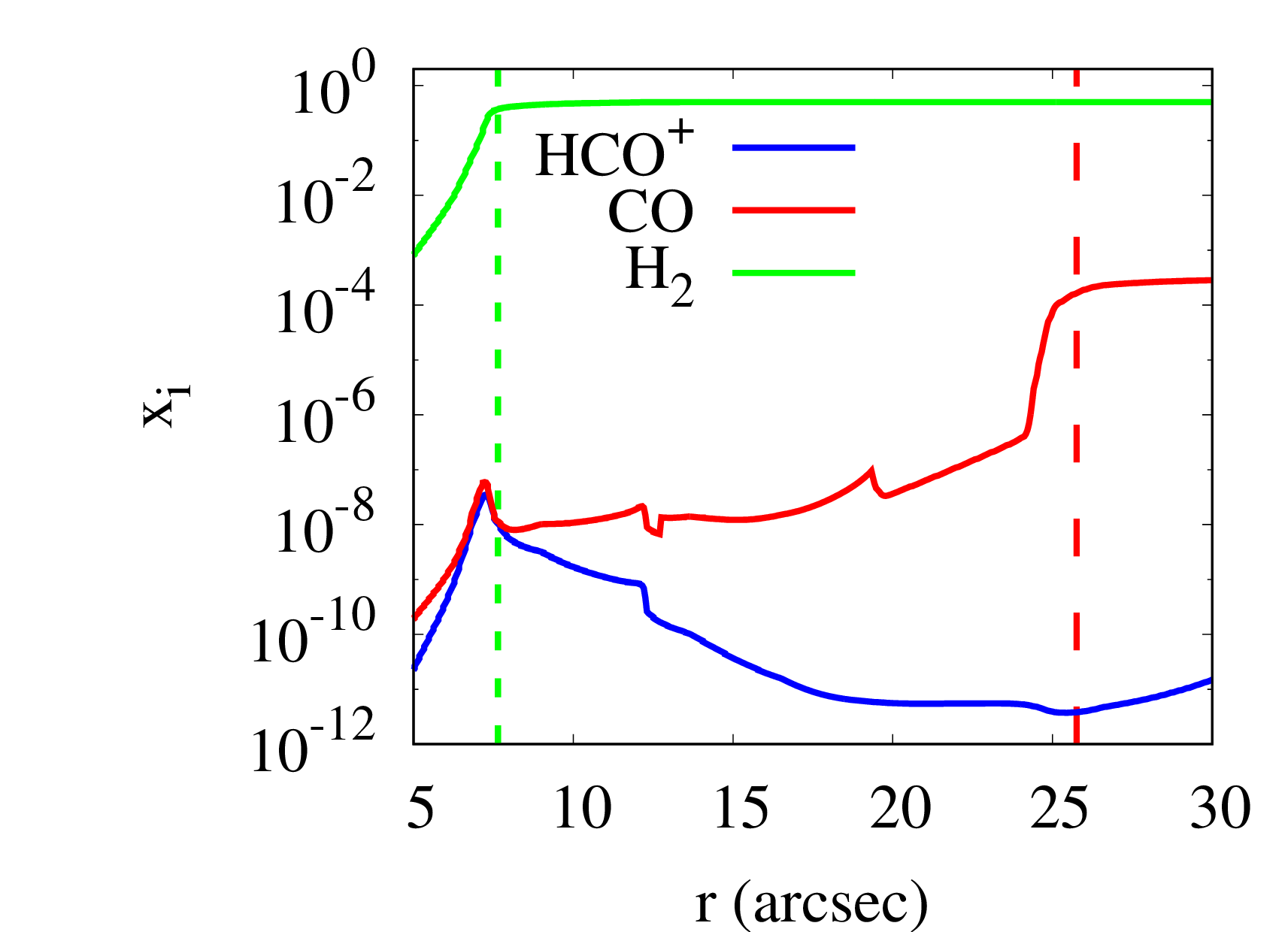}
\includegraphics[width=4.23cm,angle=270]{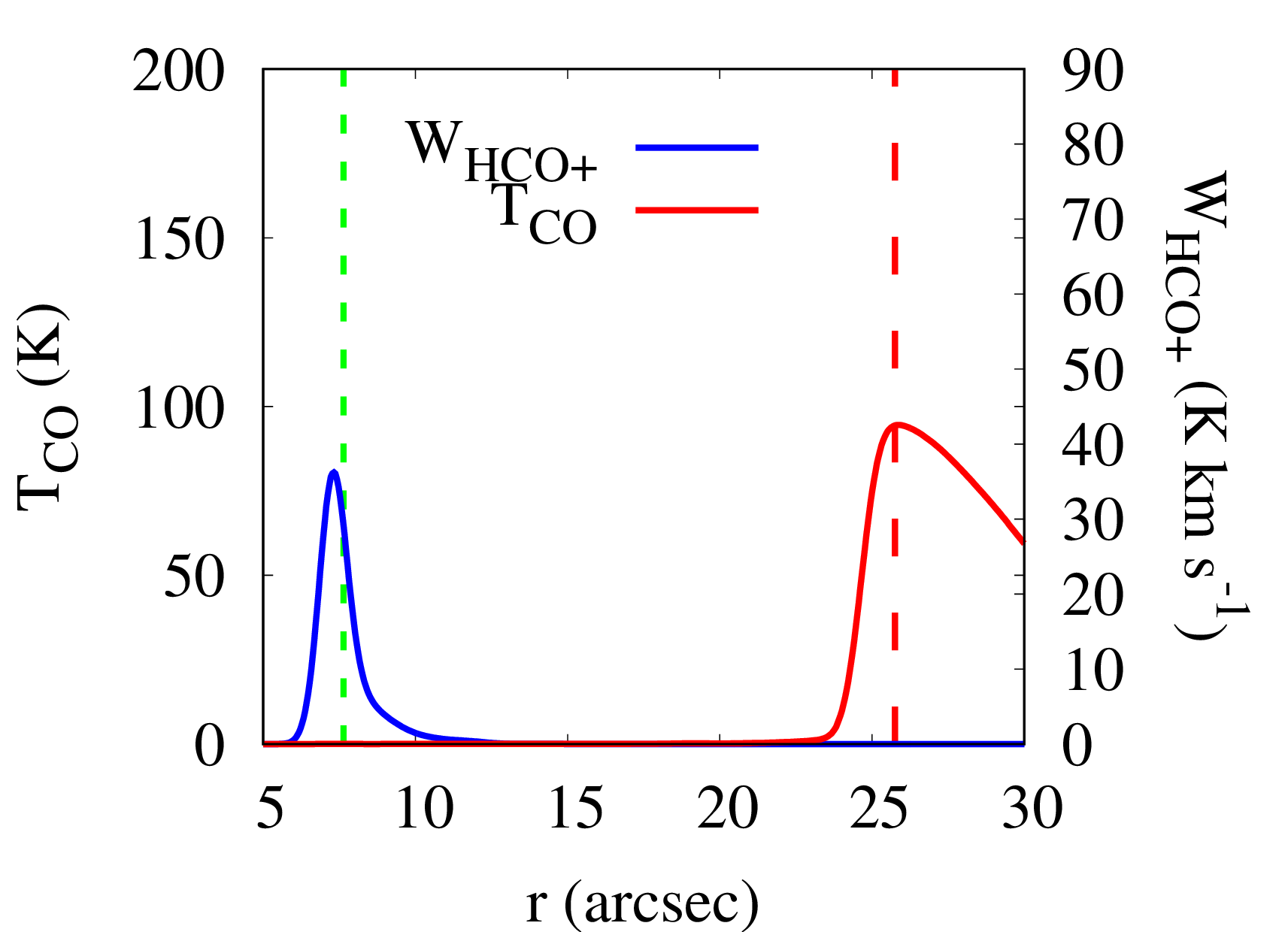}
\includegraphics[width=4.23cm,angle=270]{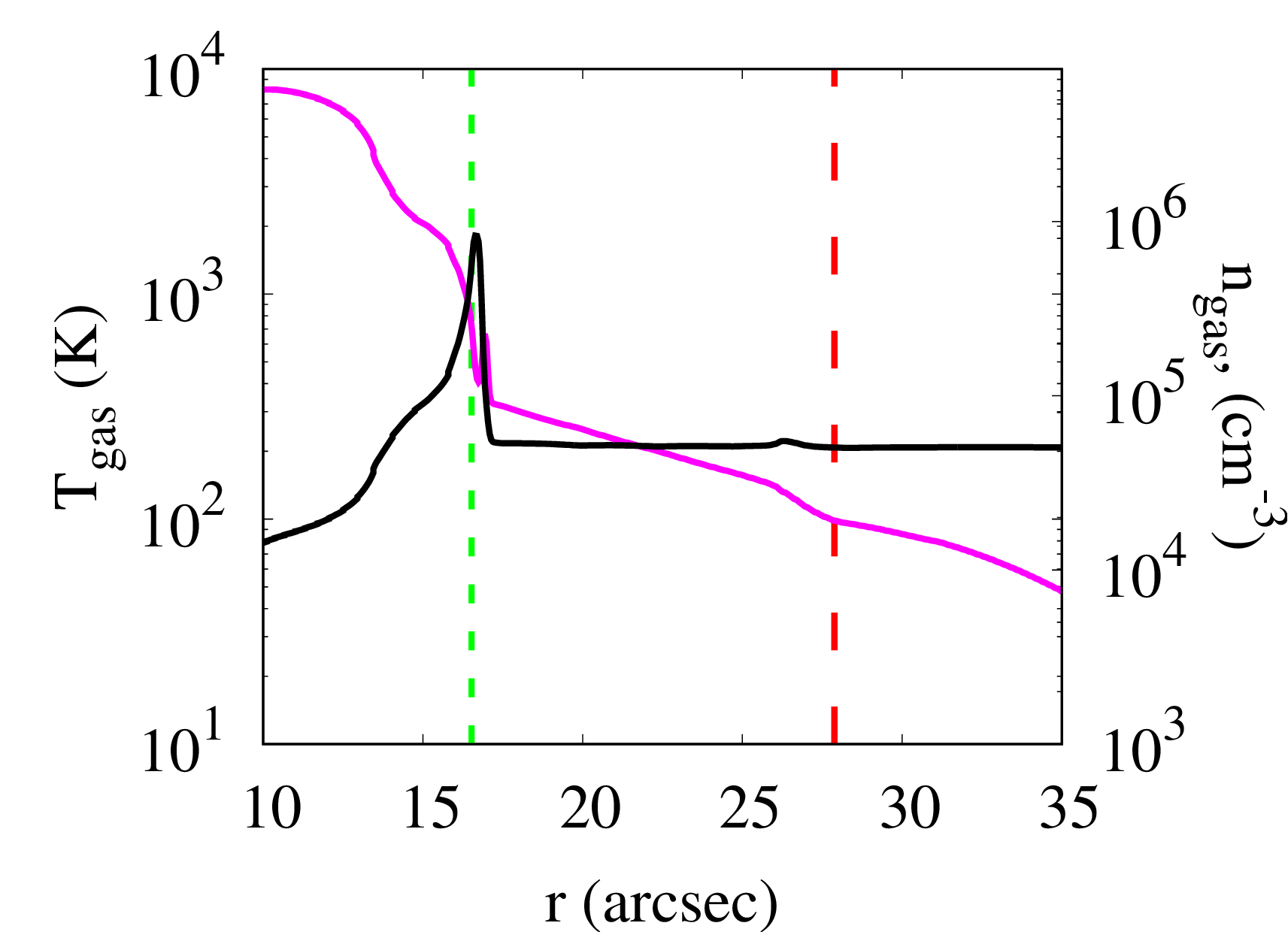}
\includegraphics[width=4.23cm,angle=270]{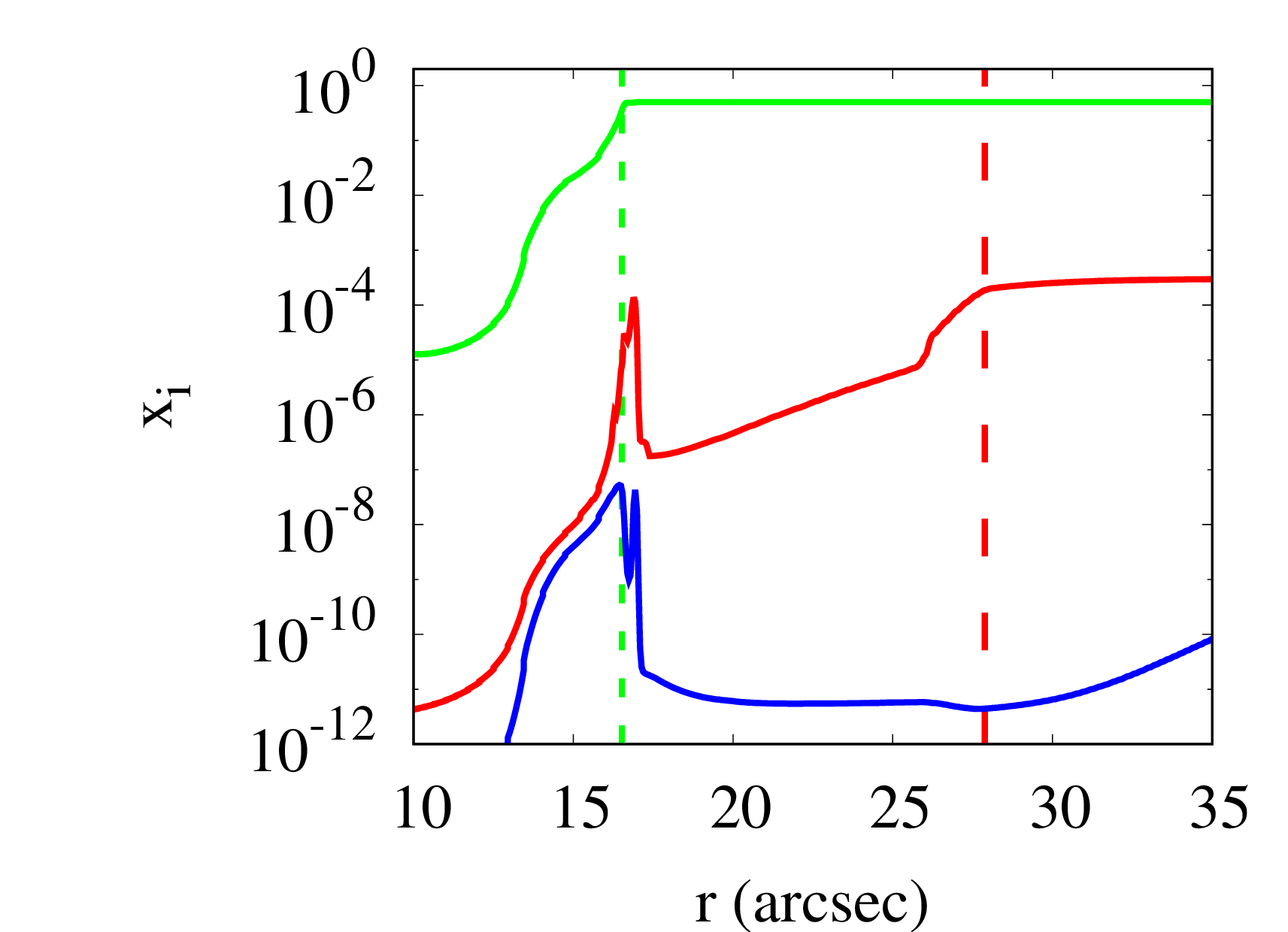}
\includegraphics[width=4.23cm,angle=270]{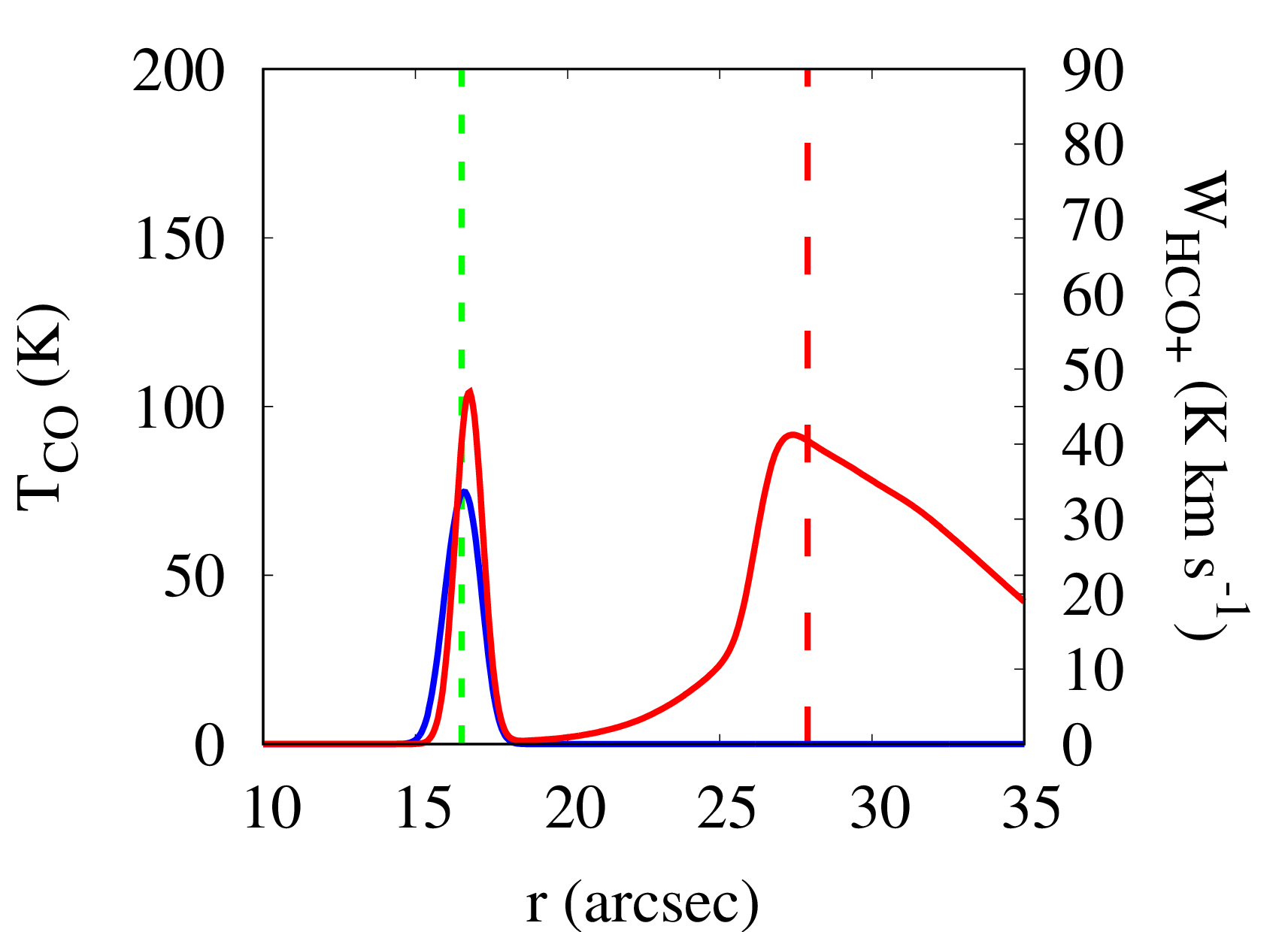}
\includegraphics[width=4.23cm,angle=270]{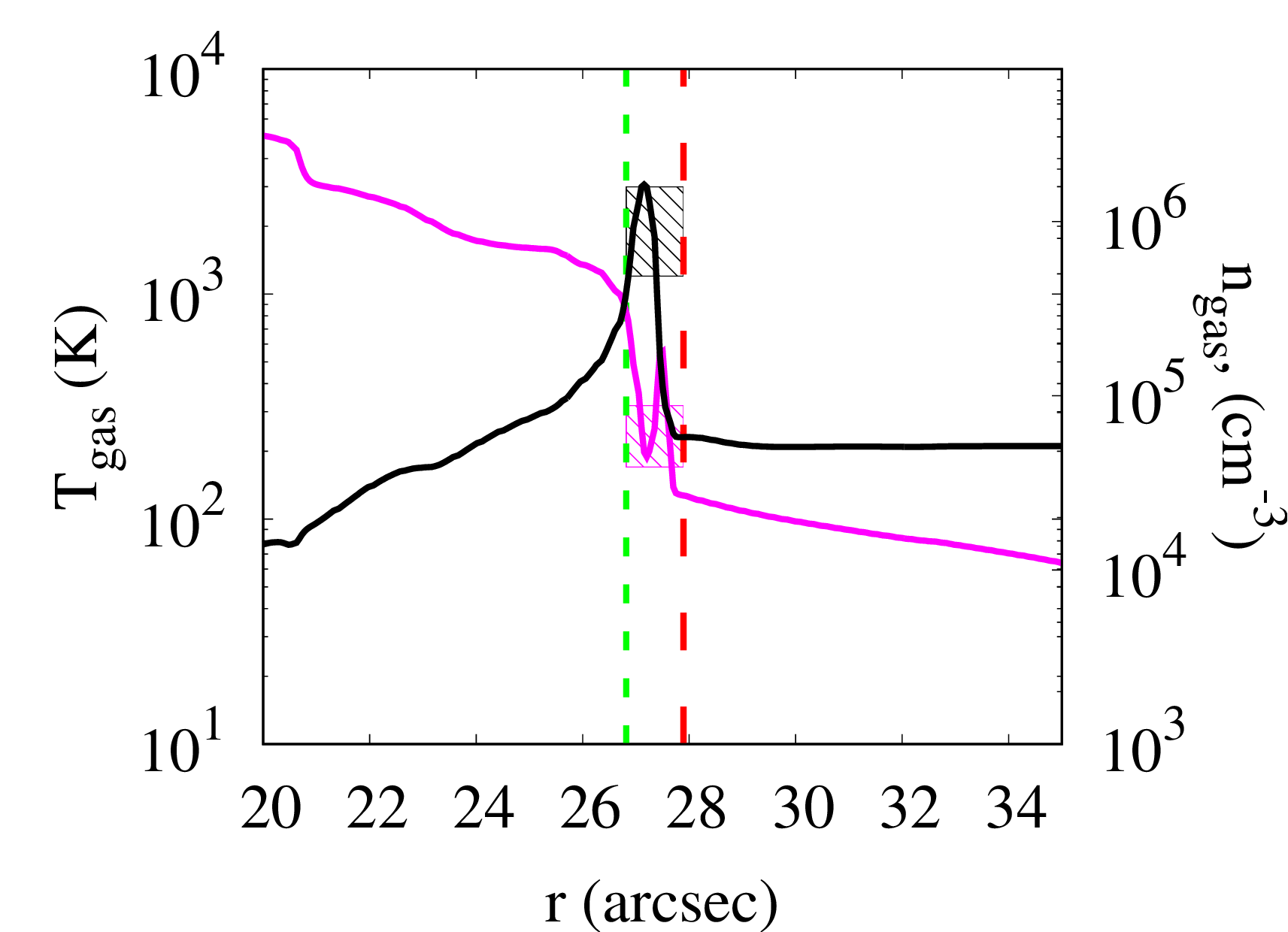}
\includegraphics[width=4.23cm,angle=270]{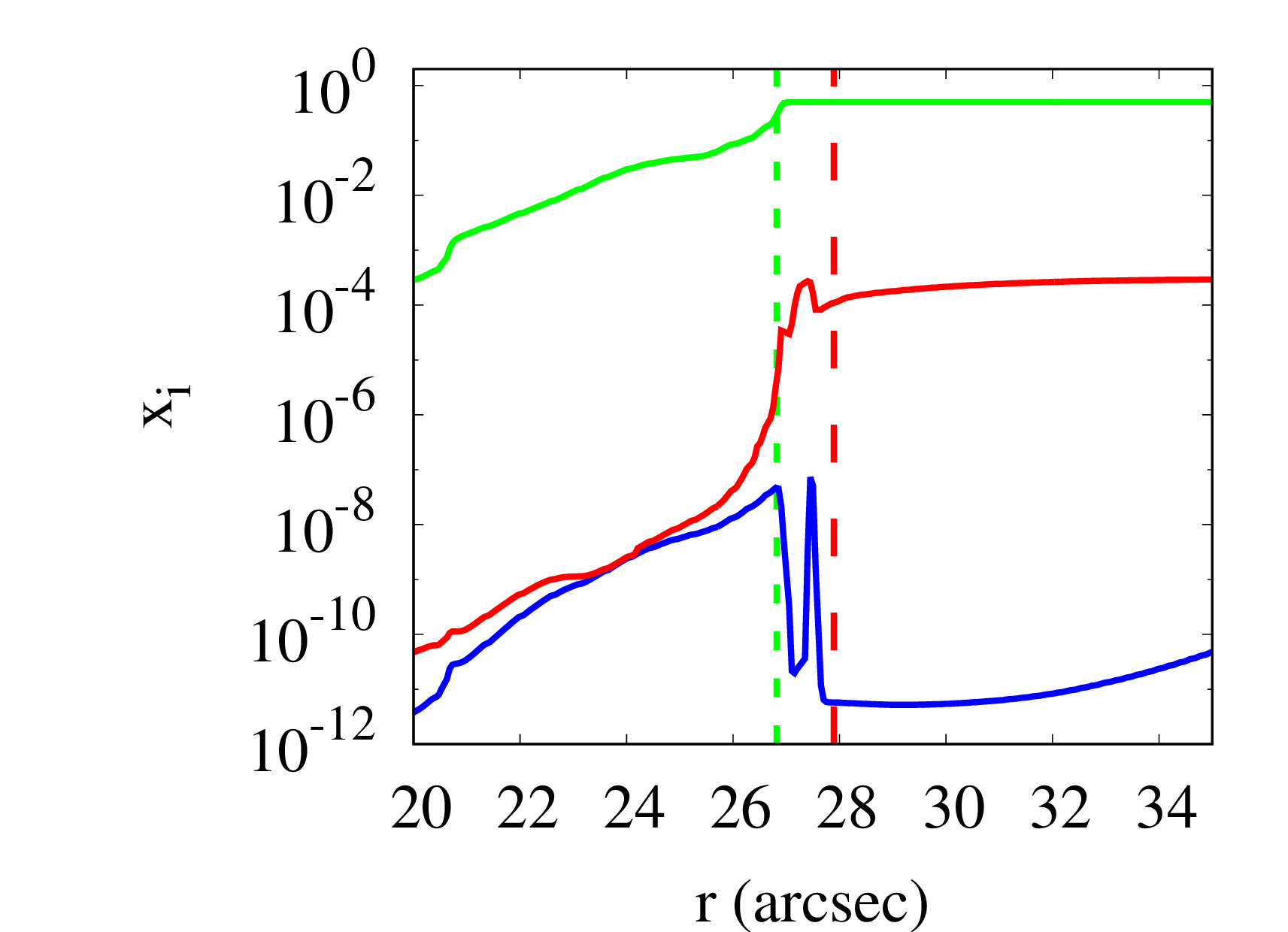}
\includegraphics[width=4.23cm,angle=270]{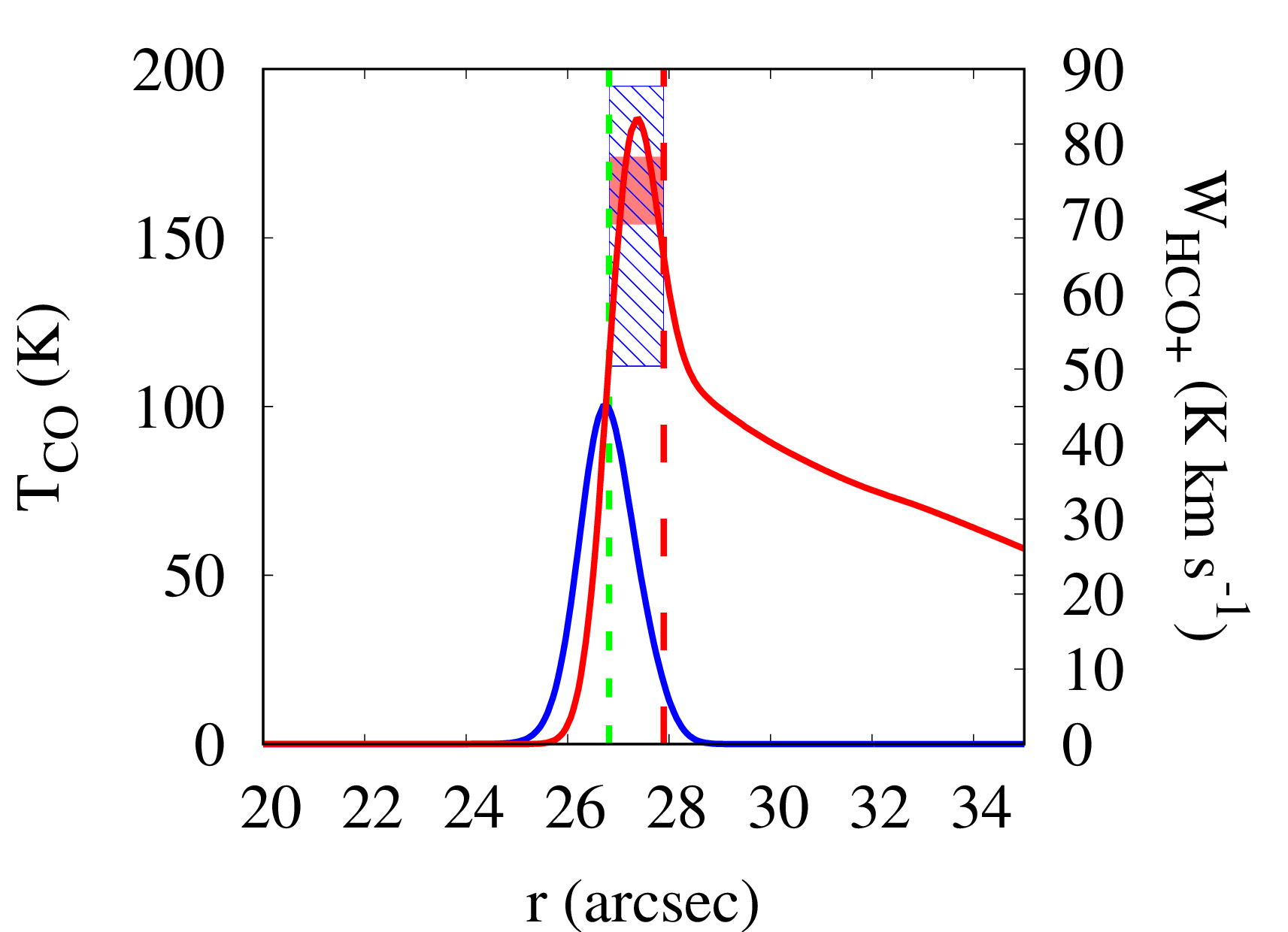}
\caption{Results of the time-dependent calculations for early ($t=2.3$\,kyr, top row), intermediate ($t=5.0$~kyr, middle row), and late ($t=8.3$~kyr, bottom row) time moments. The horizontal axis shows the distance $r$ from the origin of the computational domain. The source of UV photons is supposed to be located to the left of the computational domain. Left column: theoretical spatial distributions of \tgas\ and \ngas. Hatched rectangles on the bottom left panel (late time moment) show \tgas\ and \ngas\ values with the error bars obtained by \citet{Goicoechea_2016} using the equilibrium PDR model. Middle column: spatial distribution of H$_2$, CO and HCO$^+$ relative abundances. Right column: peak intensity of the \co\ emission line and integrated intensity of the \hcop\ emission line. Hatched rectangles with error bars indicate observational values obtained by \citet{Goicoechea_2016}. Red and green dashed lines show theoretical locations of the CO and H$_2$ dissociation fronts, respectively.}
\label{fig:timefronts}
\end{figure*}

In this section we present results of a time-dependent PDR model for a cloud that is initially completely molecular (see Table~\ref{tab:modparam}) and then instantly becomes irradiated by strong FUV field. Radiation heats up gas and drives a compression wave into the cloud. At the same time, radiation pressure pushes dust grains into the cloud, and the momentum transfer from dust to gas significantly affects the gas dynamics. FUV quanta dissociate molecules, and both H$_2$ and CO dissociation fronts also move through the cloud. We stop the calculation when the H$_2$ dissociation front crosses the outer border of the computational domain that occurs after about 13 kyr. Fig.~\ref{fig:timefronts} shows physical, chemical, and emission properties of the model cloud for three representative time moments, which correspond to 2.3~kyr, 5.0~kyr and 8.3~kyr since the beginning of the calculation and are further referred to as early, intermediate, and late time moments. The moments show three varieties of mutual locations of H$_2$ and CO dissociation fronts, mutual locations of \co\ and \hcop\ emission peaks, and their intensities.

At the early time moment gas density distribution in the considered region is still nearly flat, but a density enhancement starts to appear in the leftmost part of the region. This is where the H$_2$ dissociation front is located. The CO dissociation front initially is located somewhat deeper into the cloud; the separation between the H$_2$ and CO dissociation fronts is $\approx18$\,arcsec. Transition from H to H$_2$ occurs closer to the border of the cloud because of effective self-shielding of H$_2$ molecules against photo-dissociation. While some radiation capable of dissociating CO molecules is absorbed by H$_2$ molecules (mutual shielding), photons dissociating CO can penetrate deeper into the cloud with the nearly flat density distribution. This is why the CO dissociation front forms far ahead of the H$_2$ dissociation front at this time \citep[see][]{vanDishoeck_1988}. It is interesting that abundance of HCO$^+$ is slightly larger than CO abundance in the leftmost part of the computational domain, between the \hcop\ peak and the H$_2$ dissociation front. The theoretical \hcop\ emission is much fainter at this moment than the observed emission.

As the structure of the transition region between the cloud and the PDR evolves, \hcop\ emission brightens in the direction of the H$_2$ dissociation front, as shown in the middle row of Fig.~\ref{fig:timefronts}. The \hcop\ emission peak nearly coincides with the gas density enhancement that continues to grow ahead of the H$_2$ dissociation front. The separation between the H$_2$ and CO dissociation fronts is about 11\,arcsec, which is still larger than ALMA resolution at 350~GHz. Maximum \hcop\ integrated intensity is $\approx 70$~K km s$^{-1}$, which is in agreement with the average value observed in the Orion Bar. \co\ peak emission is seen towards the CO dissociation front in the gas that is heated by radiation but is still mostly undisturbed. There is also a secondary \co\ emission peak at $r = 17$\,arcsec that is nearly coincident with the gas density enhancement and the \hcop\ emission peak. Later this minor \co\ emission peak will merge with the bulk of the CO-emitting gas.

\begin{figure}
\centering
\includegraphics[width=5.5cm,angle=270]{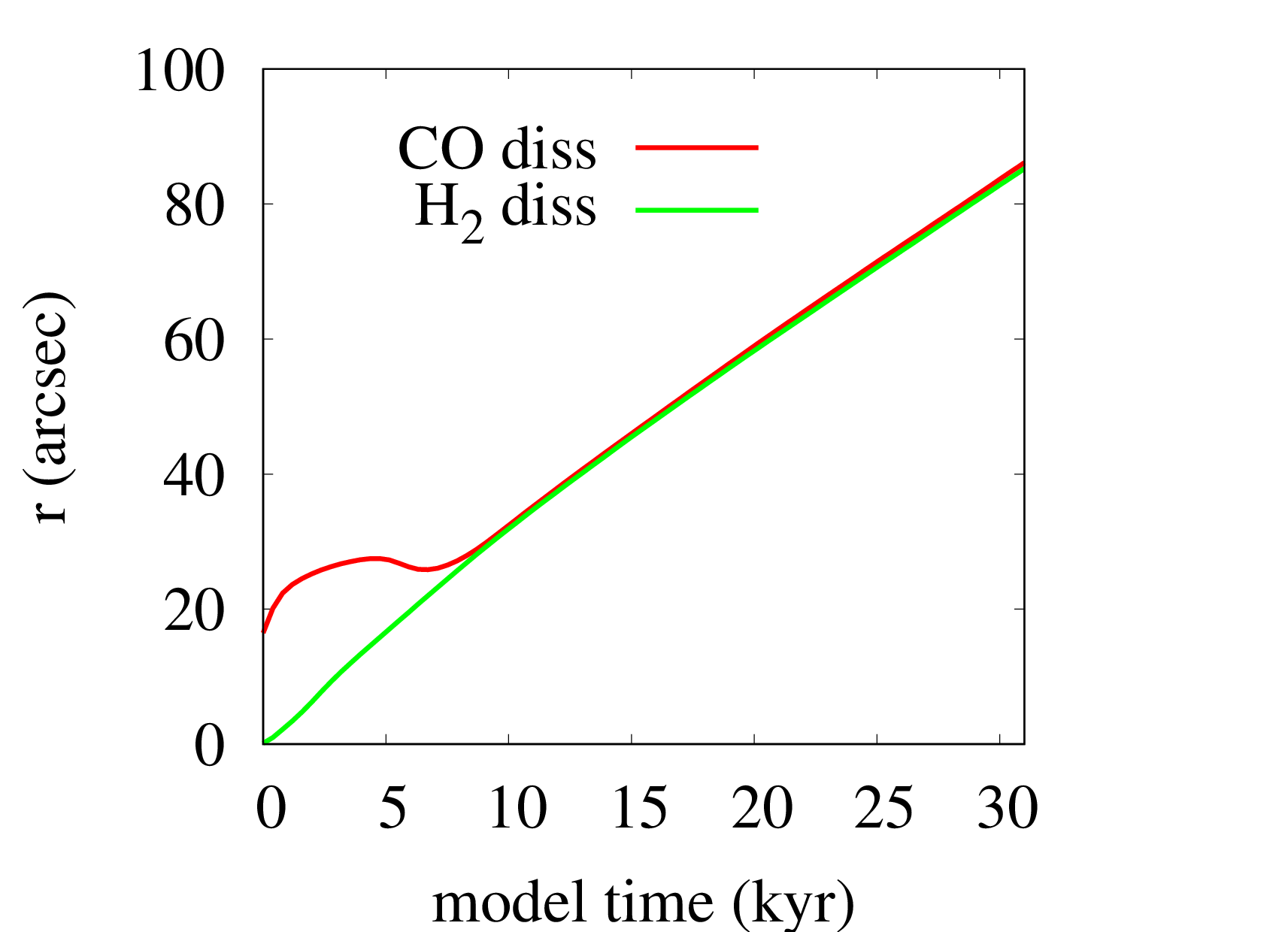}
\caption{Locations of the H$_2$ and CO dissociation fronts as a function of time in terms of distance.}
\label{frontpos}
\end{figure}

After about 7--8~kyr of evolution (the late time moment), the structure of the transition region is finally established. Its main feature is that the H$_2$ and CO dissociation fronts are nearly merged with each other (see the bottom rows in Fig.~\ref{fig:timefronts} and Fig.~\ref{frontpos}). The distance between the fronts is only 1.0--1.5\,arcsec. Since this moment, the merged H/H$_2$ and C$^+$/C/CO transitions move together away from the source of FUV radiation. 

\begin{figure}
\centering
\includegraphics[width=5.5cm,angle=270]{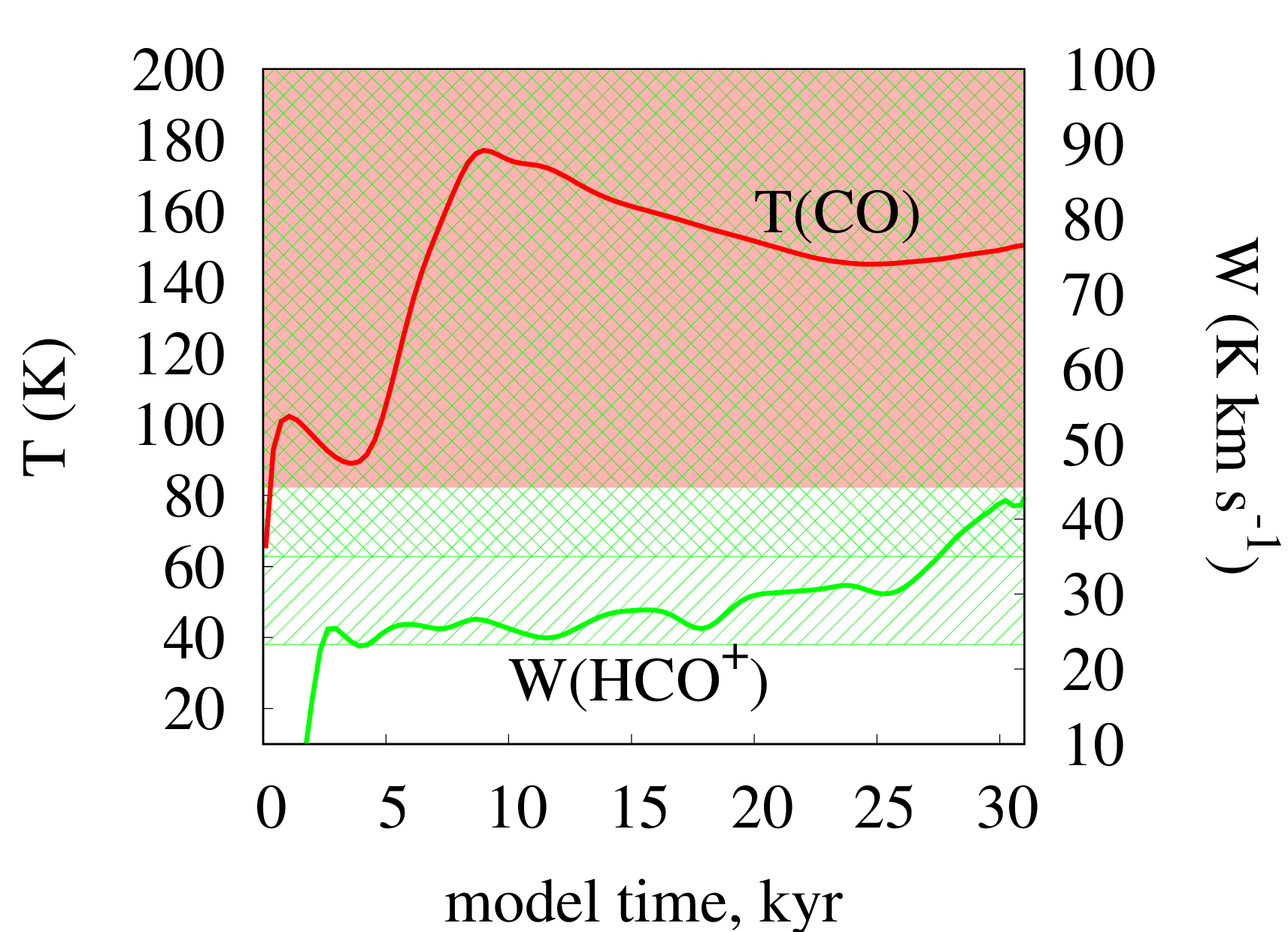}
\caption{The peak intensity of the \co\ line and the maximum integrated intensity of the \hcop\ line in the compressed layer as a function of time. Shaded and cross-hatched areas indicate intensities that differ from the observed values by no more than a factor of 2. A hatched area indicates intensities that differ from the observed values by no more than a factor of three.}
\label{brighist}
\end{figure}

The separation between the \hcop\ and \co\ emission peaks is also about 1\,arcsec after the structure is established. This value is close to ALMA resolution at 350~GHz and is, thus, consistent with the ALMA observations. The \ngas\ enhancement exceeds an order of magnitude at this moment. The density and temperature values in the compressed layer are roughly in agreement with estimates obtained by \citet{Goicoechea_2016} using the equilibrium PDR model and \hcop\ and \co\ emission analysis (in Fig.~\ref{fig:timefronts} these estimates are shown with hatched rectangles). Similar gas number density and temperature in the PDR were obtained by \citet{Sorochenko_2000} after the analysis of carbon radio recombination lines and infrared fine-structure lines of C\,{\sc{ii}} and O\,{\sc{i}} (\ngas=$(1.3\pm0.4)\cdot 10^5$~cm$^{-3}$, \tgas=$215\pm32$~K). Both the peak intensity of the \co\ line and the maximum integrated intensity of the \hcop\ line are consistent with the corresponding observed values averaged through the front in the Orion Bar. This consistency is preserved thereafter within a factor of three as illustrated in Fig.~\ref{brighist}.

The theoretical \co\ peak intensity does not differ from the observed value by more than a factor of two over the entire considered time span. This is because the peak intensity of an optically thick transition is close to the gas temperature in the molecular region, and this temperature does not vary much at the CO dissociation front and beyond. The same result is also obtained in the stationary run. Thus, the \co\ emission is indeed a reliable gas temperature measure in the Orion Bar, as suggested by \citet{Goicoechea_2016}. 

The situation with the peak of \hcop\ emission is less trivial. At the intermediate and late time moments, it nearly coincides with the peak of the gas number density. This is consistent with the suggestion made by \citet{Goicoechea_2016} on the correspondence of the \hcop\ emission peak to the gas density peak. The integrated intensity of the \hcop\ line persistently grows through the early and the intermediate times, reaching the value of about 100\,K at the late time moment and then decreases down to about 30\,K. There is a significant scatter in the computed parameters of the HCO$^+$ line, because they are very sensitive to the environmental conditions (a smoothed curve is shown in Fig.~\ref{brighist}). One can see in the bottom row of Fig.~\ref{fig:timefronts} that the peak of the HCO$^+$ emission is located at the illuminated side of the CO dissociation front. Evidently, HCO$^+$ molecules bind some carbon atoms freed from photo-dissociated CO molecules. The density peak is located in the region right next to the CO dissociation front, where HCO$^+$ abundance sharply decreases with distance. Even minor variations in the combination of HCO$^+$ abundance and the total gas density may cause large scatter in the emission parameters. We note that the chemical network we use for this study is reduced and adopted to reproduce the locations of CO and H$_2$ dissociation fronts in the equilibrium calculations \citep{Rollig_2007}. But it appears to be good enough to reproduce HCO$^+$ abundances as well. We note that the \hcop\ emission has optical depth greater than one.

\begin{figure}
\centering
\includegraphics[width=0.7\columnwidth,angle=270]{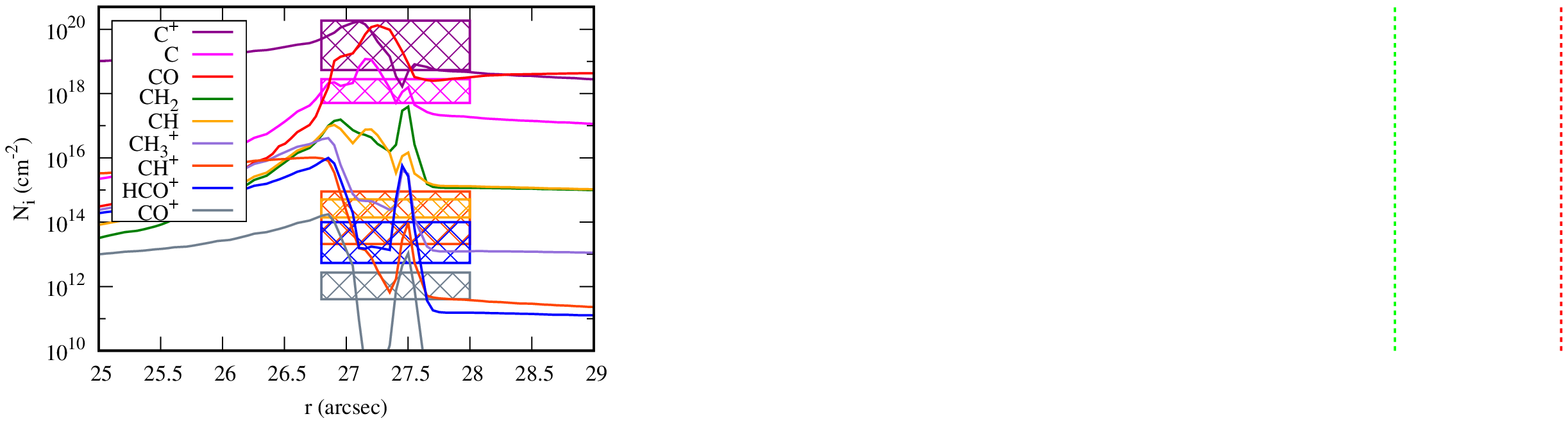}
\caption{Column densities of carbon-bearing molecules at the late time moment. }
\label{carbon}
\end{figure}

In Fig.~\ref{carbon} we show column densities of the species which contain most carbon atoms after the merging of the fronts. Various components of the carbon hydrogenation chain becomes abundant, with the most abundant CH$_2$, CH, CH$_3^+$ and CH$^+$. In Fig.~\ref{carbon} we also compare our theoretical results with observed abundances of several species in the Orion Bar measured by \citet{Nagy_2013,Nagy_2017} toward the `CO$^+$\,peak' with the {\it Herschel} telescope. Spatial resolution of the {\it Herschel} data is worse than the ALMA data discussed here. So, the abundances from {\it Herschel} are averaged over much larger area than the horizontal axis range we show in Fig.~\ref{carbon}. It is seen that we reproduce abundances of atomic carbon, C$^+$, CH$^+$, HCO$^+$ within an order of magnitude at least between the H$_2$ and CO dissociation fronts. We also reproduce abundance of CH, which has an uncertainty range much smaller (about a quarter of magnitude) than the other mentioned molecules, at least around the CO dissociation front.

We also found that the ratio of C/CO relative abundances in the PDR are in general agreement with 0.05--0.1 values reported by \citet{White_1995}. The ratio rises from 0.01 at the H$_2$ dissociation front to 0.1 at the CO dissociation front. We consider the abundance of CO$^+$ between the H$_2$ and the CO dissociation fronts as being in agreement with \citet{Stoerzer_1995, Fuente_2003}, who determined the abundance using a single-dish telescope.

The used chemical network does not contain hydrocarbon molecules with more than one carbon atom as well as species containing both carbon and nitrogen. Such a restriction of the network does not liberate the carbon atoms from the chemical chains leading to formation of CO, HCO$^+$ and CH$_4$. This is probably why we obtain high abundances of CH$_2$ and CH$_3^+$ which are intermediaries of the chemical chain for the CH$_4$ formation. Richer chemical network is needed to model the detailed  PDR chemistry which is beyond the scope of the present study.

The conclusion of this section is the following. In a dynamically evolving PDR, the H$_2$ and CO dissociation fronts rapidly merge and move together thereafter. By the end of the computation run a thin compressed gas layer forms, which is no more than 5\,arcsec wide (0.01~pc), and the thickness of the topmost density peak is about 2\,arcsec (0.005~pc). CO and HCO$^+$ emission radial profiles look like sharp peaks. Their widths and respective intensities are similar to what is observed in the Orion Bar. The observed bright \hcop\ emission is only matched by the theoretical one after the dissociation fronts merge and propagate into the surrounding molecular gas.

To explore the importance of the dynamics we made an additional test. The physical structure (gas and dust number densities) were taken as the initial conditions for the stationary calculations similar to the one presented in Section~\ref{sec:res1}. Thermal structure for gas and dust distributions was calculated from the pressure equilibrium condition. We found that no bright \hcop\ emission appears in the shocked gas at such conditions because of low \tgas\ in the region with high \ngas.

\section{Discussion}\label{sec:Disc}

Our goal is to verify whether the observed closeness of the H$_2$ and CO dissociation fronts in the Orion Bar is related to the dynamical evolution of the object. Indeed, in the previous sections we showed that the near coincidence of \co\ and \hcop\ emission peaks observed in the Orion Bar is not reproduced in the stationary PDR model and appears naturally in the dynamical model. Merging of the emission peaks and the corresponding dissociation fronts should not be considered as a certain temporary feature of the presented simulation. While we do perform time-dependent calculations, the model time cannot be interpreted as an `age' of the Orion Bar. First, we do not know when the PDR has been formed. Second, the Orion Nebula is a blister-type \hii\ region, so that the 1D MARION model cannot be applied to this object `as is'. Our results should rather be interpreted as a `stationary dynamic situation', which can only be realized when we take into account the action of pressure onto the molecular gas.

This result is consistent with conclusions that have been reached in some previous studies. For example, \citet{Hosokawa_2005} presented hydrodynamic simulations of expanding \hii\ regions that included calculation of the H$_2$ and CO dissociation. They showed that the dissociation fronts of H$_2$ and CO become closer to each other during the expansion and nearly merge in 400 kyr after the beginning of the expansion for physical conditions pertaining to \hii\ region Sh2-104. The calculation of \citet{Hosokawa_2005} is the first chemo-dynamical simulation of an expanding \hii\ region with a real hydrodynamic model. We should also mention a study of \citet{Storzer_1998} who used analytic approach to show that in non-equilibrium calculations the C$^+$/CO transition layer becomes closer to the H$_2$ dissociation front compared to the equilibrium model. They also concluded that non-equilibrium effects in the Orion Bar are probably small. This conclusion was in agreement with contemporary observational results, however, now we may argue that dynamical effects are important in highly FUV-irradiated gas.

\begin{figure*}
\centering
\includegraphics[width=5.5cm,angle=270]{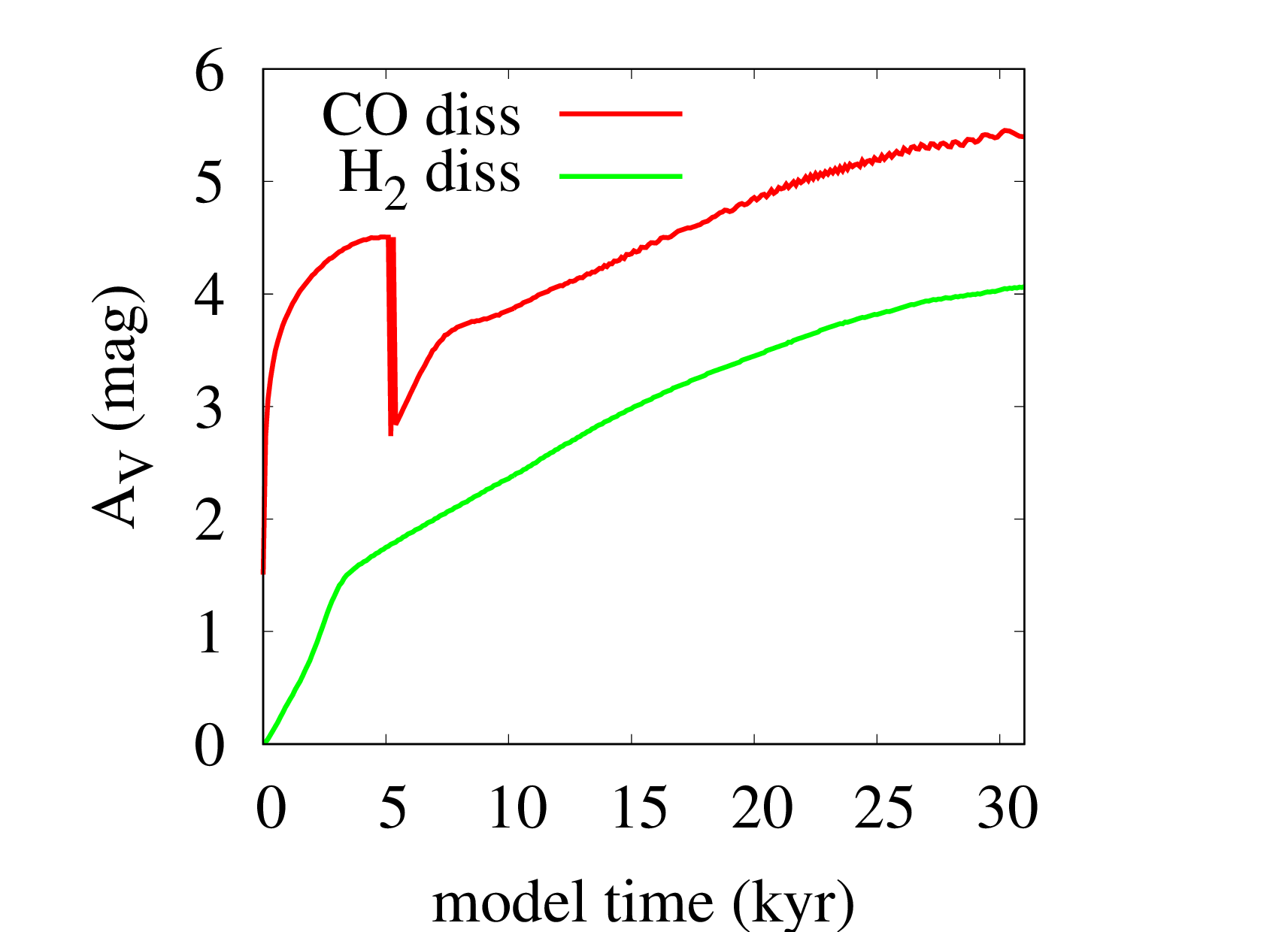}
\includegraphics[width=5.5cm,angle=270]{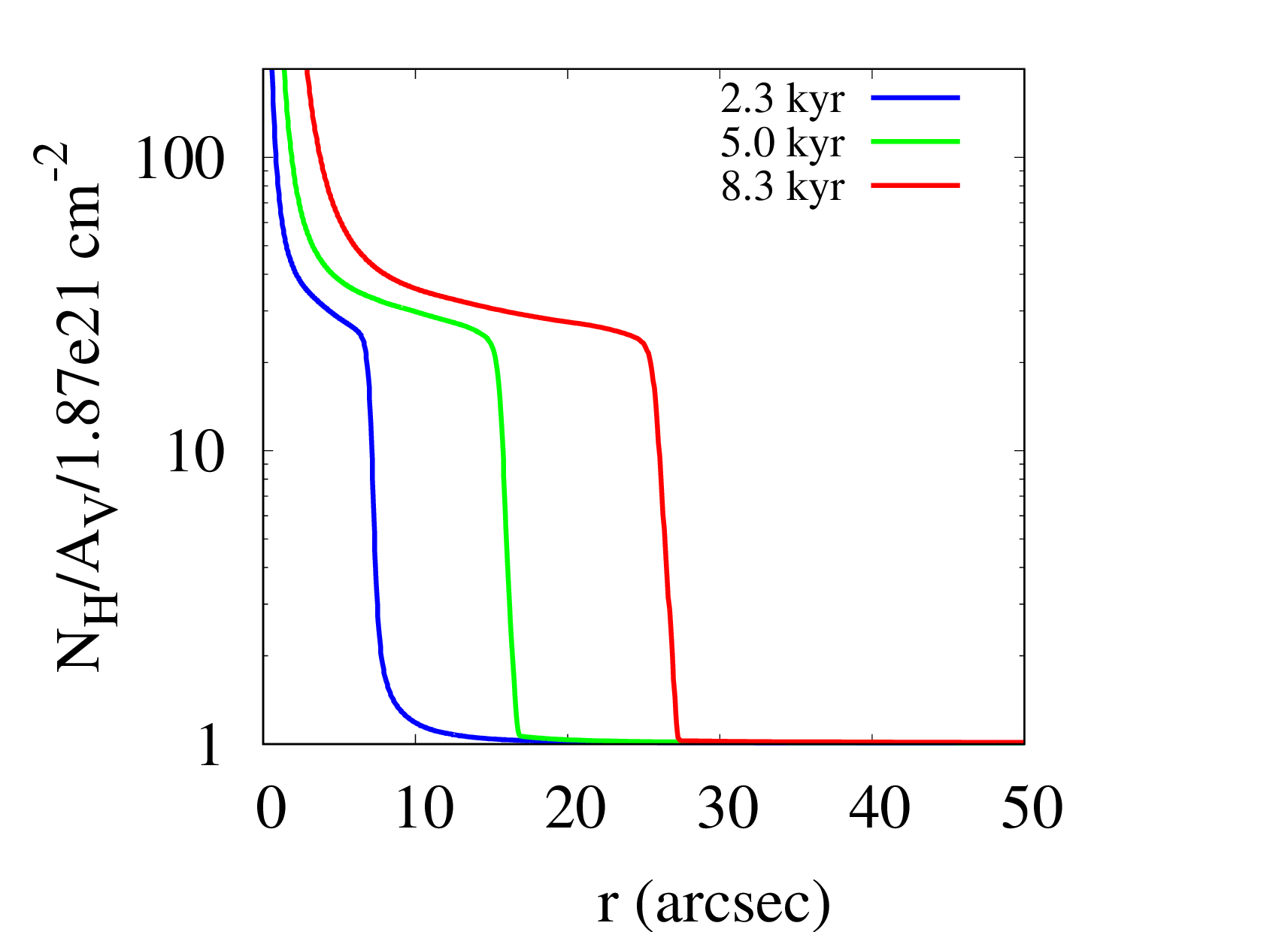}
\caption{Left: locations of the H$_2$ and CO dissociation fronts as functions of time in terms of $\av$ value. Right: relation between column density of hydrogen atoms and $\av$ value for the same time moments as shown in Fig.~\ref{fig:timefronts}.}
\label{fronts}
\end{figure*}

Studies of PDRs commonly predict that the HI/H$_2$ transition occurs at $\av\approx2$\,mag, and the C$^+$/C/CO transition occurs at $\av\approx4$\,mag \citep{Tielens_1985,1994ApJ...422..136T}. Our results are generally consistent with this picture as seen in the left panel of Fig.~\ref{fronts}, where $A_V=\tau_{V}/1.086$, and $\tau_{V}$ is computed for the current dust distribution. The diagram shows locations of the H$_2$ and CO dissociation fronts in terms of $\av$. It is seen that after the stationary dynamic situation has been established on a time-scale of about 7--8~kyr, which is significantly smaller than the typical age of an expanding \hii\ region around an O-type star \citep[e.g.][]{spitzerbook}, the fronts moves slowly from $\av$ = 3\,mag and 4\,mag for H$_2$ and CO, respectively, to $\av$ = 4\,mag (H$_2$) and 5.5\,mag (CO). The front convergence is caused by the appearance of the density enhancement. Because of the sharp density increase in the enhancement the transition region between the illuminated medium to the shaded medium becomes quite narrow, and all the major chemical transformations occur over a very limited distance range. This result is naturally absent in stationary models that assume constant density across the PDR. One may argue that the nearly coincident dissociation fronts can be naturally reproduced by a constant pressure model with a non-uniform density distribution. However, our tests indicate that it is impossible in this case to reproduce bright \hcop\ emission.

\citet{Joblin_2018} claim the existence of thin (few 10$^{-3}$~pc corresponding to $0.5-1$\,arcsec) emission layers at high thermal pressures about 10$^8$~K\,cm$^{-3}$ in the Orion~Bar PDR. Our computed physical conditions in the compressed gas layer for the late time moment presented in the Fig.~\ref{fig:timefronts} agree with their suggestion. It is noteworthy that the similar value of the thermal pressure in the compressed layer was obtained by \citet{Bron_2018} using photoevaporation flow model. Edge-on orientation of the Orion Bar PDR does not allow testing gas kinematics, although studies of this kind could possibly give us a direct answer about the dynamical model, appropriate for the region. The Orion Bar is irradiated not only by O7 $\theta^1$C~Ori star from north-west but also by O9.5 $\theta^2$A~Ori~\citep{2011ApJS..193...24S} from south-east. This other star may also influence the chemical structure of the Orion Bar. While $\theta^1$C~Ori dominate ionization in the inner part of the Orion Nebula \citep{ODell_2017}, we still believe that influence of $\theta^2$A~Ori on the molecular abundances towards the south-east edge of the Bar should be further investigated.

It is worthwhile to relate our results to those presented by \citet{Pellegrini_2009} and \citet{Shaw_2009}. These authors have explored the Orion Bar observations, presented by \citet{Tielens_1993} and \citet{1994ApJ...422..136T} for a different position, where H$_2$ and CO(1--0) emission peaks appear to be separated by an angular distance of about 10 arcsec. \citet{Pellegrini_2009} and \citet{Shaw_2009} have been able to reproduce both the location and the intensity of the emission peaks for the [\sii], H$\alpha$, H$_2$, and CO(1--0) lines, using a stationary model with the enhanced magnetic field and CR ionization rate. The value of the CR ionization rate had to be increased by nearly four orders of magnitude in order to reproduce observed CO and H$_2$ intensities in simulations of \citet{Pellegrini_2009} and \citet{Shaw_2009}.

The ALMA observations and our simulations draw a somewhat different picture, in which H$_2$, \co{} and \hcop{} emission peaks nearly coincide, without enhancing the cosmic ray ionization rate above the standard ISM value. Even more so, raising the ionization rate by just one order of magnitude may lead to some serious chemical consequences. We tested higher values of the CR ionization rate up to $5\times10^{-16}$\,sec$^{-1}$, considering recent results of \citet{Sorochenko_2010, Indriolo_2012, Neufeld_2017}. The peak values of the \co\ and \hcop\ emission intensity in the PDR remain the same as in our basic model. The difference, significant for the present study, is the high HCO$^+$ abundance in the undisturbed molecular region, where the abundance is defined by reactions with H$_3^+$ ion.

These different results may be somehow related to the non-uniform matter distribution within the Orion Bar. \citet{Goicoechea_2016} and \citet{AndreeLabsch_17} demonstrate a clumpiness of the Orion Bar. So, some disagreement between our results and the observed picture can be caused by the fine structure of the PDR unresolved in our model. Also, our restricted simulations do not include modelling of the Orion Nebula's ionization front so we can not relate mutual positions of the ionization and dissociation fronts. In conclusion we emphasize that the model suggested by \citet{Pellegrini_2009} and \citet{Shaw_2009} successfully describes spatially resolved observations in optics and infrared while our model reproduces the spatially resolved observations in millimeter wavelengths. In this sense, these two models compliment each other. A more advanced modelling that includes both dynamics and chemistry,  considers the ionized, atomic and molecular gas, and takes into account more molecular line data \citep[e.g.,][]{2018A&A...617A..77P,2019A&A...622A..91G}, will definitely help to resolve the issue.


Fig.~\ref{fronts} (left) shows that a value of $\av$ corresponding to the location of the CO dissociation front drops by one magnitude at the late time moment when the H$_2$ and CO dissociation fronts merge at the border of the PDR. This drop reflects a fact that dust grains and gas can move relative to each other in the used version of the MARION model. This is further illustrated in the right panel of Fig.~\ref{fronts}, which shows how relation between the hydrogen column density and visual extinction changes in the PDR and the molecular cloud. Due to dust motion through the gas and variable dust-to-gas mass ratio, we see different $N_{\rm H}/A_{\rm V}$ ratios. Within the molecular cloud this ratio is the same as in the original dust model. At the density peak and within the PDR, dust is depleted due to the drift driven by radiation pressure. The important difference is that at the early and intermediate time moments the CO dissociation front is located farther into the cloud, far ahead of the dust depletion zone, while at the late time moment and afterwards the CO dissociation front is located within the dust depletion zone. This is why $A_{\rm V}$ is lower at the front location. In the absence of the abundant dust, the CO protection from photodissociation is mostly provided by H$_2$ shielding. Thus, CO survival is directly related to H$_2$ survival. This effect further narrows the gap between the two dissociation fronts. The enhanced $N_{\rm H} / A_{\rm V}$ ratio found in the present study is in agreement with previous findings by \citet{Abel_2004,Abel_2006,Abel_2016} in the Orion Veil, which is physically associated with the Trapezium stars and the Orion Nebula.


To strengthen this conclusion, we repeated the calculation assuming that dust is frozen to gas. In this case, the evolution of the PDR including the front convergence proceeds somewhat slower. The separation between the fronts becomes smaller than the ALMA resolution after about 13 kyr, which is still far less than other relevant time-scales of the PDR evolution.

\citet{Salgado_2016} report the decrease of dust opacity in the Orion~Bar~PDR. Specifically, they find a factor of four and five decrease of infrared and ultraviolet dust opacities, respectively, in the ionized gas compared to the value for diffuse interstellar gas. Factor of five decrease of the infrared dust opacity is found towards the PDR itself. These results support our conclusion on the enhanced $N_{\rm H} / A_V$ ratio in the Orion~Bar~PDR. \citet{Salgado_2016} obtain the increase of the dust optical depth at 19.7~$\mu$m from 0.025 in the ionized gas to 0.2 in the PDR. These values are also in agreement with the variation of $\tau_{19~\mu{\rm m}}$ between the H$_2$ and CO dissociation fronts obtained with the presented MARION model.

Some important remarks should be made before we conclude. First, we tried various initial conditions for the calculations. Specifically, we tested initial densities in the range of $5\times 10^{4}-10^{5}$~cm$^{-3}$ and $\chi$ in the range of $10^{4} - 10^{5}$. We also considered the Orion Veil $A_{\rm V}/N_{\rm H}=3.3\times10^{-22}$~cm$^2$ ratio for the calculations of the Orion Bar PDR~\citep{Abel_2004,Abel_2006,Abel_2016} and found no difference in terms of the \co\ and \hcop\ emission intensity in the PDR with our basic model. The mentioned test calculations bring us to a conclusion that radial profiles of various molecular lines, similar to those presented by \citet{Goicoechea_2016} for CO and HCO$^+$, and the abundance profiles are highly desirable to testify the predictions of PDR models and confine the physical parameters. Ever for such an extensively studied region as the Orion Nebula and its surroundings we did not find the abundance profiles in the literature to compare with the predictions from the Fig.~\ref{carbon}.


\section{Conclusion}\label{sec:conc}

We argue that dynamical effects in the FUV-irradiated gas, namely, the formation of the compressed layer at the border of the PDR and dust drift driven by the radiation pressure, are responsible for the observational appearance of the Orion Bar on the \co\ and \hcop\ maps at 350~GHz produced by ALMA. The interaction between the heated PDR and the cold molecular cloud leads to the formation of the density enhancement that slowly moves into the cloud. As the mutual location of the transition regions depends on $A_{\rm V}$, and its value grows on a short distance scale within the compressed layer, the regions become close to each other, so that their angular separation is too small to be resolved with the ALMA observations. Another important factor that speeds up the formation of the observed configuration is the dust expulsion from the PDR by the stellar radiation pressure. The resultant chemical and thermal structure leads to the correspondingly coincident locations of the H$_2$, \co\ and \hcop\ emission peaks. Finally, high gas number density and high relative abundance of HCO$^+$ cause bright \hcop\ line emission, which appears on the illuminated side of the CO dissociation front. These observed features are not reproduced by the stationary model and appear naturally in dynamical model of the PDR. We also confirm suggestions of \citet{Goicoechea_2016} that \hcop\ emission corresponds the gas density peak in the compressed region and that \co\ line emission is a reliable measure of the gas temperature in the Orion Bar.

\section*{Acknowledgments}\label{sec_ackn}

We are thankful to Ya. N. Pavlyuchenkov for fruitful discussions of the radiation transfer with RADEX. The anonymous referee is gratefully acknowledged for his/her comments and suggestions which allows us to improve the paper.

The study is supported by RFBR grant number 16-02-00834-a.

\bibliographystyle{mnras}
\bibliography{Oriondyn-ac}

\appendix
\section{Comparison with other PDR codes} \label{AppA}

\begin{figure*}
\includegraphics[width=16cm]{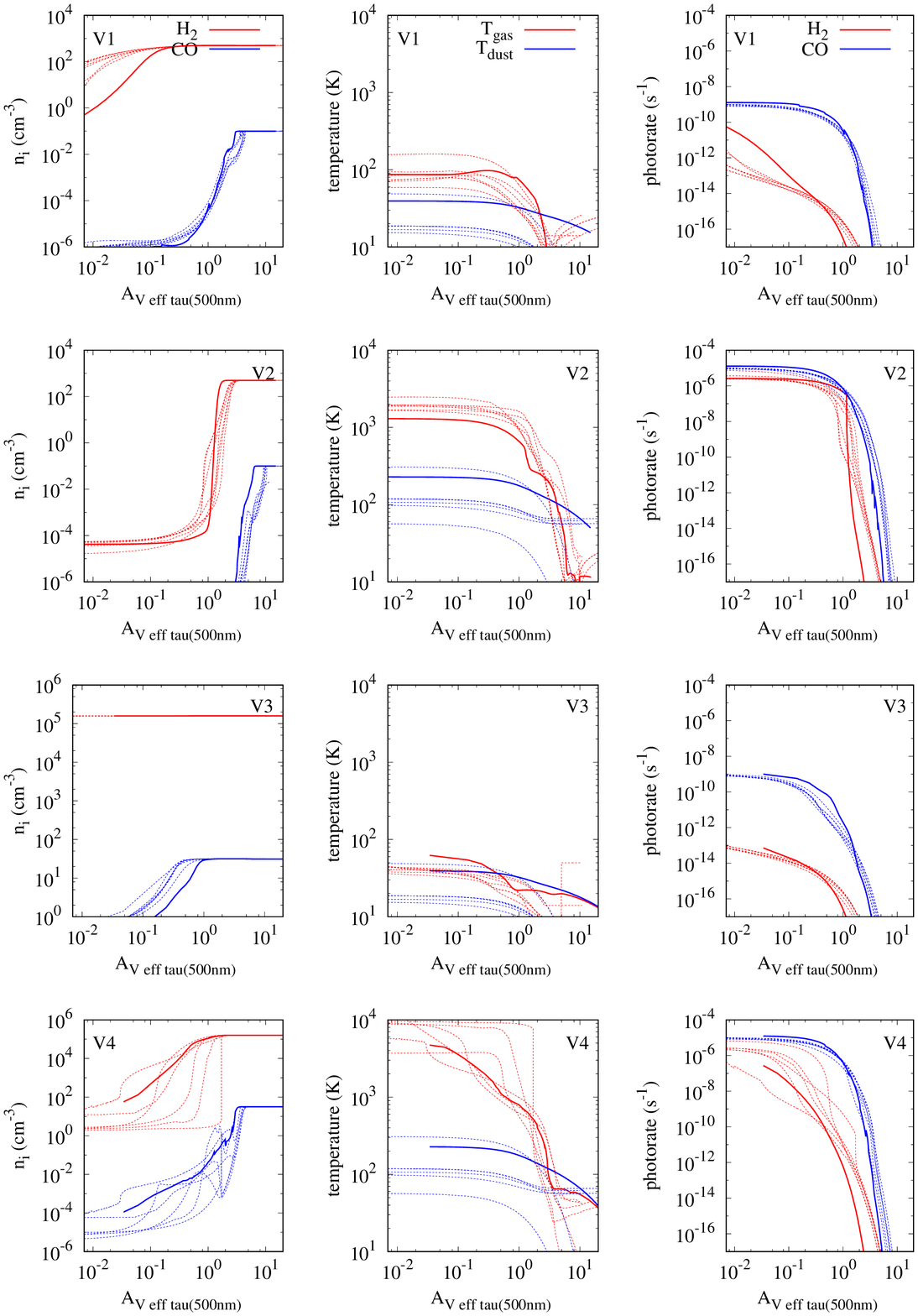}
\caption{Comparison of MARION calculations with the results of the PDR benchmark meeting. The MARION results are shown with a solid line, while dashed lines show other PDR models. Left column: number densities of H$_2$ and CO. Middle column: gas and dust temperatures. Right column: rates of H$_2$ and CO photodissociation.}
\label{fig:pdrbench}
\end{figure*}

We performed tests of the MARION model using results of the PDR-benchmarking workshop held in 2004 \citep{Rollig_2007}. All F1-F4 (fixed gas and dust temperature) and V1-V4 (variable gas and dust temperature) models are tested. We show here comparison of the MARION results with \citet{Rollig_2007} for V1-V4 models. In Fig.~\ref{fig:pdrbench} results of MARION are shown by a solid line and results of Meudon, HTBKW, KOSMA-$\tau$, Leiden, Meijerink and Sternberg are shown by dashed lines. All the references to the codes and initial conditions for the test calculations are given in \citet{Rollig_2007}.

\bsp	
\label{lastpage}
\end{document}